\documentclass[pra,twocolumn,amsmath,amsfonts,amssymb,a4paper,floatfix]{revtex4}
\usepackage{amsthm}
\usepackage{graphicx}

\newcommand{\bbar}[1]{\Bar{\Bar{#1}}}
\newcommand{\bm}{\mathbf}

\begin{document}

\title{Electromagnetic fields and boundary conditions at the interface of generalized transformation media}
\author{Luzi Bergamin}
\email{Luzi.Bergamin@tkk.fi}
\affiliation{European Space Agency, The Advanced Concepts Team (DG-PI), Keplerlaan 1, 2201 AZ Noordwijk, The Netherlands}
\date{December 14, 2009}

\begin{abstract}
In this paper, the interface between two transformation media or between a transformation medium and vacuum is studied.  Strictly from the transformation optics point of view, the consequences of the boundary conditions at such interfaces are addressed in two different ways. First, we analyze a restricted class of reflectionless interfaces, for which the tools of transformation optics allow to describe the electromagnetic fields on both sides of the interface by means of the same vacuum solution of the Maxwell equations. In a second step, we examine interfaces between two arbitrary transformation media. This analysis is extended to the recently suggested generalization of transformation optics by the author. As a basic application it is shown how the standard law of reflection and refraction at an interface between vacuum and a homogeneous and isotropic medium with arbitrary and \emph{independent} permittivity and permeability can be understood in a completely geometric way by the use of generalized transformation optics.
\end{abstract}

%\pacs{42.70.-a}

\maketitle

\section{Introduction}
\label{sec:intro}
Transformation optics \cite{Pendry:2006Sc,Leonhardt:2006Nj,Leonhardt:2008Oe} in the recent years became an important design tool for new types of artificial materials (metamaterials.) In transformation optics one starts from the constitutive equations of vacuum\footnote{Often the terms constitutive equation or constitutive relation refer to media, only. Since in the present context the constitutive relations of transformation media are closely related to the vacuum relation, we follow Refs.~\cite{Post,Leonhardt:2006Nj,Bergamin:2008Pa} and call the following equation constitutive relation of vacuum.} of a possibly curved spacetime, written in generic coordinates \footnote{Throughout this paper all equations are written in natural units, $\epsilon_0 = \mu_0 = c = 1$. Furthermore Einstein's summation convention is used, which implies a summation over all repeated indices. Further details of our notation are explained in Appendix \ref{sec:conventions}.}:
\begin{align}
\label{epsorig}
 D^i &= \frac{g^{ij}}{\sqrt{-g_{00}}} E_j - \frac{g_{0j}}{g_{00}} \epsilon^{jil} H_l \\
\label{muorig}
 B^i &= \frac{g^{ij}}{\sqrt{-g_{00}}} H_j + \frac{g_{0j}}{g_{00}} \epsilon^{jil} E_l
\end{align}
It is then observed that these equations resemble the constitutive equations of a special medium with $\epsilon^{ij} = \mu^{ij} = g^{ij}/\sqrt{-g_{00}}$ and bi-anisotropic contributions $\xi^{ij} = - \zeta^{ij} = - g_{0l} \epsilon^{lij}/g_{00}$ \cite{Landau2}. Thus, if empty spacetime can look like a medium, it should be possible to find media that look like empty spacetime. Transformation media \cite{Leonhardt:2006Sc,Pendry:2006Sc,Leonhardt:2006Nj} are media of this type. They are linear media that may be interpreted as to mimic a different spacetime. In other words, the solutions of the Maxwell equations in the medium, which is placed in a certain spacetime called laboratory space, can be mapped on the solutions of the electromagnetic fields propagating in a different, but empty spacetime. Transformation optics thus is a tool to express the solution of the Maxwell equations in a yet unexplored medium (the transformation medium) in terms of well-known solutions (here vacuum solutions.) Additionally, transformation optics allows to design media with a pre-defined propagation of light (a pre-defined solution of Fermat's principle \cite{Leonhardt:2008Oe}) in an easy way, since this propagation is encoded geometrically in the chosen transformation of spacetime, locally expressed as a coordinate transformation. As most popular examples, light can be guided around a volume in space (e.g.\ a sphere or a cylinder) leading to an invisibility cloak \cite{Pendry:2006Sc,Leonhardt:2006Sc} or the transformation medium can mimic an inversion of space, which leads to a perfect lens \cite{Leonhardt:2006Nj}. In these applications it appeared natural to make the interface between vacuum (outer space) and the transformation medium reflectionless. Though transformation optics today is used in a much broader context than these two examples, somewhat surprisingly the complete conditions for a reflectionless interface only have been presented recently \cite{Yan:2008Rl} and the study of boundary conditions at a generic interface still seems to be missing.

Despite its successes, transformation optics also has its limitations, mainly in terms of the accessible effective media parameters. As can be seen from Eqs.\ \eqref{epsorig} and \eqref{muorig}, the constitutive relation of vacuum always has the form of a reciprocal medium with permittivity and permeability being equal. These restrictions led to ideas how the original setup could be generalized, either within the geometric approach of transformation optics \cite{Bergamin:2008Pa} or by replacing geometric transformations by direct field transformations \cite{Tretyakov:2008Gf}. Though these generalizations also provide a mapping of vacuum solutions of the Maxwell equations onto the solutions of the medium, the implications of this map are often less immediate than in standard transformation optics, where the medium just mimics a free spacetime. Thus a thorough analysis of reflection and refraction at interfaces between such media or between a medium of this type and vacuum is important to improve our understanding of these tools.

It is the aim of this paper to study interfaces between two arbitrary transformation media in detail. We will concentrate on standard transformation media or media of the generalization of Ref.~\cite{Bergamin:2008Pa}, which will be explained more in detail in Sect.~\ref{sec:troptics}. At these interfaces standard boundary conditions,
\begin{align}
\label{BC1}
 (\bm D_1 - \bm D_2 )\cdot \bm n &= -\sigma\ , & (\bm B_1 - \bm B_2 )\cdot \bm n &= 0\ , \\
 \label{BC2}
 (\bm E_1 - \bm E_2) \times \bm n &= 0\ , & (\bm H_1 - \bm H_2) \times \bm n &= \bm K\ ,
\end{align}
will be imposed ($\bm n$ is the unit vector normal to the interface and the indices 1 and 2 refer to the two different sides of the interface.) The implications of these boundary conditions can be studied in two different ways. Since a solution of the Maxwell equations in a transformation medium is expressed in terms of a vacuum solution, one can ask the question for which combinations of media the boundary conditions are met if the same vacuum solution is used on both sides of the interface. In this approach (worked out in detail in Sect.~\ref{sec:boundaryI}) the implementation of the boundary conditions yields constraints on the geometric transformations used to describe the two media and it is shown that these constraints can be reduced to two simple rules. They provide a sufficient (though not necessary) condition for a reflectionless interface. If these constraints are met everywhere on the interface, this interface disappears completely in the formulation in terms of vacuum solutions and consequently becomes invisible.

Of course, one can consider the interface between two arbitrary transformation media, which in general is not reflectionless. In this case, as discussed in Sect.~\ref{sec:boundaryII}, the solutions in the two different media are expressed in terms of two different vacuum solutions. Still, it is possible to re-formulate the boundary conditions completely in terms of the vacuum solutions and the geometric manipulations. In this reformulation the boundary conditions are no longer independent of the media as is the case in Eqs.\ \eqref{BC1} and \eqref{BC2}, but depend on the geometric transformations and thus on the characteristics of the media (the physical  content of the boundary conditions of course remains unchanged.) As discussed in Sect.\ \ref{sec:conclusions} one advantage of this formulation is the fact that the implications of the boundary conditions for a whole class of media derive from only one specific formula. As basic examples we will show in Sect.\ \ref{sec:boundaryII} how the laws of reflection and refraction at the interface between vacuum and a homogeneous and isotropic medium can be understood in a geometric way.

\section{Generalized transformation optics}
\label{sec:troptics}

In this section a brief introduction to standard transformation optics and a generalization thereof are presented. Originally, transformation optics was introduced as a tool to design invisibility cloaks \cite{Pendry:2006Sc,Leonhardt:2006Sc}, the full concept as used in this paper was introduced in Ref.~\cite{Leonhardt:2006Nj}, a pedagogical review was presented by the same authors in Ref.~\cite{Leonhardt:2008Oe}.

\begin{figure}[t]
 \centering
 \includegraphics[width=\linewidth,bb=0 0 256 43]{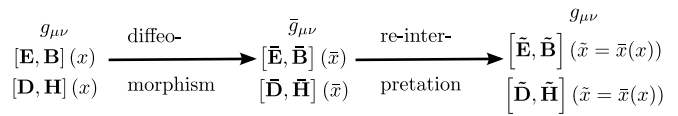}
 % leonhardt.eps: 0x0 pixel, 300dpi, 0.00x0.00 cm, bb=0 0 256 43
 \caption{Illustration and notation for standard transformation optics.}
 \label{fig:leonhardt}
\end{figure}

As already mentioned in the introduction, transformation optics is based on the fact that the constitutive relations of vacuum of a possibly curved spacetime and written in general coordinates resemble those of a reciprocal medium. Thus it should be possible to find media, which may be interpreted as to mimic an empty spacetime, which however is different from the spacetime the medium is placed in. Though in principle no restrictions on the nature of the spacetime to be mimicked exist, the two spacetimes in most applications are related by a diffeomorphism, locally implemented as a coordinate transformation (see Fig.~\ref{fig:leonhardt}.)

To design a specific medium by means of transformation optics one starts with the definition of laboratory space, which is the spacetime where the transformation medium shall be placed in. In this spacetime a coordinate system with coordinates $x^{\mu}$ is chosen and in these coordinates the spacetime metric $g_{\mu\nu}(x)$ takes a certain form. One can write down the vacuum solutions of the Maxwell equations in laboratory space, which we denote by $\bm E(x)$, $\bm B(x)$, $\bm D(x)$ and $\bm H(x)$. Now a mapping from laboratory space to a different spacetime, called electromagnetic space, is defined. Mathematically this transformation is a diffeomorphism, locally it is implemented as a coordinate transformation $x^\mu \rightarrow \bar x^\mu (x)$. Since electrodynamics is invariant under diffeomorphisms, the two spacetimes are physically equivalent and one simply rewrites the vacuum solutions in terms of the new coordinates $\bar x^{\mu}$ and the new metric $\bar g_{\mu\nu}$. Until now no medium parameters have been defined, but the Maxwell equations just have been rewritten in terms of different coordinates. In a second step, it is claimed that the physical spacetime still is laboratory space, but the electromagnetic fields shall propagate as if the spacetime was electromagnetic space. This makes the presence of a medium necessary. Technically this means that the solutions $\bar{\bm E}(\bar x)$, $\bar{\bm B}(\bar x)$, $\bar{\bm D}(\bar x)$ and $\bar{\bm H}(\bar x)$, which are solutions of the Maxwell equations in the spacetime with metric $\bar g_{\mu\nu}$, have to be turned back into solutions in the spacetime with metric $g_{\mu\nu}$. Since the Maxwell equations in general coordinates only depend on the determinant of the spacetime metric (see Eqs.~\eqref{EOMcomp} and \eqref{covder}), this can be achieved by a simple rescaling of fields. As has been shown in Refs.~\cite{Leonhardt:2006Nj,Bergamin:2008Pa}, a possible rescaling is
\begin{align}
\label{scal1}
 \tilde E_i &= \bar s \bar E_i\ , & \tilde B^i &= \bar \sigma \frac{\sqrt{\bar \gamma}}{\sqrt{\gamma}} \bar B^i\ , \\
 \label{scal2}
 \tilde D^i &= \bar \sigma \frac{\sqrt{-\bar g}}{\sqrt{-g}} \frac{\partial \bar x^0}{\partial x^0} \bar D^i\ , & \tilde H_i &= \bar s \frac{\sqrt{-g_{00}}}{\sqrt{-\bar g_{00}}} \frac{\partial \bar x^0}{\partial x^0} \bar H_i\ .
\end{align}
Here, $\gamma$ is the determinant of the induced space metric according to Eq.~\eqref{indmetric}, $\bar s = \pm 1$ is positive, if the transformation $x^{\mu}\rightarrow \bar x^{\mu}$ does not change the orientation of the manifold (i.e.\ a right-handed coordinate system in laboratory space is mapped onto a right-handed one in electromagnetic space) and $-1$ otherwise. Finally, $\bar \sigma = \pm 1$ is positive, if space and spacetime in electromagnetic space have the same orientation, $-1$ otherwise. The signs $\bar \sigma$ and $\bar s$ play an important role in the context of negative refractive index media \cite{Leonhardt:2006Nj} and will be written explicitly in all equations to keep full generality of the result. Since this paper does not deal specifically with negative refractive index media, these signs are not explained in detail at this point. Further comments are made in the Appendix, for a detailed discussion we refer to \cite{Bergamin:2008Pa}. Our notation is also summarized in Fig.~\ref{fig:leonhardt} and explained more in detail in the Appendix \footnote{As is seen from Fig.~\ref{fig:leonhardt}, the explicit coordinates of laboratory space with medium are distinguished from the ones of empty laboratory space. This is necessary, since a particular point in spacetime may be represented by \emph{different} values of the coordinates in empty laboratory space and in laboratory space with the medium.}.

If the barred electromagnetic fields constitute a solution of the Maxwell equations in electromagnetic space then it is easy to check that the fields with a tilde are indeed a solution in laboratory space. It is important to notice that the rescalings \eqref{scal1} and \eqref{scal2} are not a symmetry transformation and thus the barred solutions are not physically equivalent to the solutions labeled with a tilde. Instead, by means of this rescaling a medium has been introduced which mimics the electromagnetic space in laboratory space. From the constitutive relation in electromagnetic space, Eqs.~\eqref{epsorig} and \eqref{muorig} in terms of barred variables, the constitutive relation of the transformation medium is easily derived with Eqs.~\eqref{scal1} and \eqref{scal2} as
\begin{align}
\label{epstilde2}
 \tilde{ D}^i &= \bar s \frac{\bar g^{ij}}{\sqrt{-\bar g_{00}}}  \frac{\sqrt{\bar \gamma}}{\sqrt{\gamma}} \tilde E_j - \frac{\bar g_{0j}}{\bar g_{00}} \epsilon^{jil} \tilde{ H}_l\ , \\
\label{mutilde2}
 \tilde B^i &= \bar s \frac{\bar g^{ij}}{\sqrt{-\bar g_{00}}} \frac{\sqrt{\bar \gamma}}{\sqrt{\gamma}} \tilde{ H}_j + \frac{\bar g_{0j}}{\bar g_{00}} \epsilon^{jil} \tilde E_l\ .
\end{align}
We mention that transformation optics does not only provide the constitutive relation and the solutions of the Maxwell equations, but it also defines the dispersion relation\footnote{This dispersion relation is the result of the mathematical manipulations of transformation optics and it is not claimed that it corresponds to any real medium.} in a purely geometric way. The vacuum dispersion relation, $\mathbf{k}^2 = \omega^2$, rewritten in terms of the generic coordinates of the electromagnetic space becomes $\bar g^{\mu\nu} k_{\mu} k_{\nu} = 0$ with $(k_{\mu}) = (\omega, k_i)$. This relation also has to hold in the transformation medium, but now is interpreted in laboratory space.

\begin{figure}[t]
\begin{center}
 \includegraphics[width=\linewidth,bb=0 0 256 105]{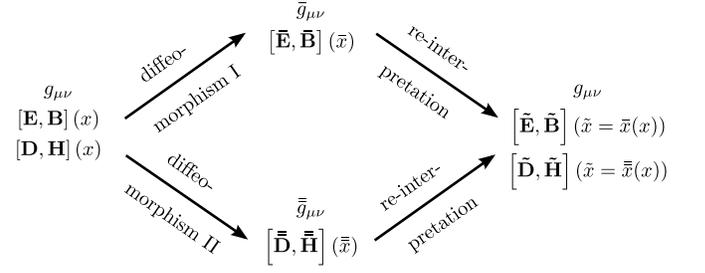}
 % triple.eps: 1179666x1179666 pixel, 300dpi, 9987.84x9987.84 cm, bb=0 0 256 105
\end{center}
 \caption{Illustration and notation of generalized transformation optics according to Ref. \cite{Bergamin:2008Pa}. Notice that the diffeomorphism I only acts on the fields $\bm E$ and $\bm B$, while diffeomorphism II acts on $\bm D$ and $\bm H$.}
 \label{fig:triplespace}
\end{figure}
In Ref.~\cite{Bergamin:2008Pa} an extension of transformation optics was introduced, which is based on the same geometrical principles as standard transformation optics but allows to design media not covered by the constitutive relations \eqref{epstilde2} and \eqref{mutilde2}. This extension starts from the observation that the Maxwell equations,
\begin{align}
\label{maxwell}
 \nabla_i B^i &= 0\ , & \nabla_0 B^i + \epsilon^{ijk}\partial_j E_k &= 0\ , \\
\label{maxwell2}
 \nabla_i  D^i &= \rho\ , & \epsilon^{ijk} \partial_j  H_k - \nabla_0  D^i &= j^i\ ,
\end{align}
split into two sets of equations with mutually excluding field content. Thus, the sets of fields $(\bm E, \bm B)$ and $(\bm D, \bm H)$ can be transformed independently \footnote{These independent transformations do not establish a  symmetry, since the constitutive relation is not invariant. Nevertheless they are invariant transformations of the equations of motion.}. This means that the generalized transformation media do not mimic a single electromagnetic space, but rather two electromagnetic spaces (see Fig.~\ref{fig:triplespace}.) In these media $\bm E$ and $\bm B$ propagate as if the spacetime had metric $\bar g_{\mu\nu}$, while $\bm D$ and $\bm H$ mimic a spacetime with metric $\bbar g_{\mu\nu}$. Still, all three spaces (laboratory space and the two electromagnetic spaces) are related by diffeomorphisms and thus are physically equivalent. As indicated in Fig.~\ref{fig:triplespace}, all variables referring to the electromagnetic space of $\bm E$ and $\bm B$ are written with a bar, while the variables of the electromagnetic space of $\bm D$ and $\bm H$ are double-barred. As in standard transformation optics the solution in laboratory space is obtained from the solution in the electromagnetic spaces by suitable rescalings of the fields. These rescalings are equivalent to Eqs.~\eqref{scal1} and \eqref{scal2} if on the right hand side of the two equations in \eqref{scal2} all barred variables are replaced by double-barred ones. Now, the most general constitutive relation of generalized transformation optics can be derived as
\begin{align}
\label{tripeleps}
 \tilde{D}^i &= - \bar s \frac{\sqrt{-\bbar g}}{\sqrt{\gamma} g_{\bar 0 \bbar{0}}} g^{\bbar{i} \bar j} \tilde E_j - \bar s \bbar{s} \frac{\sqrt{-\bar g}\sqrt{-\bbar{g}}}{\gamma g_{\bar 0 \bbar{0}}} g^{\bbar{i} \bar k} g^{\bbar{0} \bar l} \epsilon_{klm} g^{\bar m\bbar j} \tilde{ H}_j\ , \\
\label{tripelmu}
 \tilde B^i &=  - \bbar s \frac{\sqrt{-\bar g}}{\sqrt{\gamma} g_{\bar 0 \bbar{0}}} g^{\bar{i} \bbar j} \tilde{ H}_j + \bar s \bbar{s} \frac{\sqrt{-\bar g}\sqrt{-\bbar{g}}}{\gamma g_{\bar 0 \bbar{0}}}  g^{\bar{i} \bbar k} \epsilon_{klm} g^{\bbar l \bar 0} g^{\bbar m \bar j} \tilde E_j\ .
\end{align}
Here $\bar s$ and $\bbar s$ are $+1$ if the corresponding maps do not change the orientation of the manifold and $-1$ otherwise. As abbreviation the notation
\begin{equation}
\label{gbbb}
 g^{\bbar{\mu} \bar \nu} = \frac{\partial \bbar{x}^{\mu}}{\partial x^{\rho}} \frac{\partial \bar{x}^{\nu}}{\partial x^{\sigma}} g^{\rho \sigma} = \bbar{g}^{\mu \rho} \frac{\partial \bar{x}^{\nu}}{\partial \bbar x^{\rho}} = \frac{\partial \bbar{x}^{\mu}}{\partial \bar x^{\rho}} \bar g^{\rho \nu}
\end{equation}
has been introduced. It is important to realize that $g^{\bbar{\mu} \bar \nu}$ is not an (inverse) spacetime metric, in particular it needs not be a symmetric matrix and it does not necessarily have signature $(3,1)$.

The most important physical differences between the two constitutive relations \eqref{epstilde2}/\eqref{mutilde2} and \eqref{tripeleps}/\eqref{tripelmu} can be summarized as follows: In standard transformation optics permittivity and permeability are always equal and proportional to the induced spatial metric $\bar g^{ij} = \bar \gamma^{ij}$ (see Eq.~\eqref{indmetric}.) This implies that $\epsilon^{ij} = \mu^{ij}$ are symmetric matrices with three positive or three negative eigenvalues. In contrast to this result, $g^{\bbar{i} \bar j}$ in Eq.~\eqref{tripeleps} and $g^{\bar{i} \bbar j}$ in Eq.\ \eqref{tripelmu} are not spatial metrics and thus non-reciprocal media with an anti-symmetric contribution to permittivity and permeability can be described within the generalized setup. Also, the eigenvalues of $g^{\bbar{i} \bar j}$ need not all be positive, which allows to obtain media with strong anisotropy (also called indefinite media.) Though permittivity and permeability are not equal, they are still related as $\bbar{s} \sqrt{-\bbar{g}} \mu^{ij} = \bar s \sqrt{-\bar g} \epsilon^{ji}$. As an interesting special case a proportionality $\mu^{ij} = \alpha \epsilon^{ji}$ with a positive or negative proportionality constant $\alpha$ is conceivable. An example of this type is presented in Sect.~\ref{sec:boundaryII}.

Standard transformation optics intends to mimic as well as possible the electromagnetic space by means of a medium. In particular, in transformation media the trajectories of light according to Fermat's principle are equivalent to the ones in electromagnetic space (but differ from the ones in laboratory space without medium; this aspect is discussed in detail in Ref.~\cite{Leonhardt:2008Oe}.) This is also seen from the fact that the Poynting vector $S^i = \epsilon^{ijk} E_j H_k$ transforms under the spatial part of the coordinate transformations in the same way as $B^i$ and $D^i$. This does not apply to generalized transformation optics as defined in Ref.~\cite{Bergamin:2008Pa}, since $\bm E$ and $\bm H$ are not affected by the same coordinate transformation (see Fig.~\ref{fig:triplespace} and the discussion in Ref.~\cite{Bergamin:2008Pa}.)

Finally we notice that bi-anisotropic contributions to the constitutive relations require a mixing of space and time in the transformation from laboratory space to electromagnetic space, such that $\bar g_{0i} \neq 0$ and/or $\bbar g_{0i} \neq 0$. In the main part of this paper we will not consider media of this type, some comments about bi-anisotropic media are made in Sect.~\ref{sec:conclusions}.

\section{Boundary conditions for media without bi-anisotropic terms}
\label{sec:boundary}
In the main part of this work we will assume media without bi-anisotropic contributions to the constitutive relation, in other words $\bar g_{0i}$ in Eqs.\ \eqref{epstilde2}, \eqref{mutilde2} and $g_{\bar 0 \bbar i}$, $g_{\bbar 0 \bar i}$ in Eqs.\ \eqref{tripeleps}, \eqref{tripelmu} are assumed to vanish. Bi-anisotropic media in principle can be treated along the same lines, however, the specific results get much more complicated.

We want to study interfaces between two media of this type or between a medium and empty space, whereby empty space is interpreted as a trivial transformation medium (all mappings are identity maps, thus the vacuum solution in laboratory space is mapped onto itself.) At the interface we will impose the standard boundary conditions \eqref{BC1} and \eqref{BC2}. Furthermore surface charge and current will be set to zero, $\sigma = 0$ and $\bm K = 0$. Of course, different types of boundary conditions could be studied as well, which however should be rather straightforward once the formalism itself has been developed.

As mentioned in the introduction already, the consequence of the boundary conditions \eqref{BC1} and \eqref{BC2} can be studied in two different ways. Since the solutions of the Maxwell equations in the media are constructed out of vacuum solutions we can ask the question under which restrictions on the transformations the boundary conditions are met if the same vacuum solution is used on both sides of the interface. This means that the interface disappears in electromagnetic space. Since the trajectories of light rays in the media are equivalent to the ones in electromagnetic space it is evident that all interfaces of this type are reflectionless. Alternatively interfaces between two arbitrary transformation media can be considered. Since these interfaces are not necessarily reflectionless, the interface does not disappear in electromagnetic space, but will be visible as a discontinuity between two different vacuum solutions. Still, it should be possible to express the laws of reflection and refraction in terms of vacuum solutions.

In both cases the exact knowledge of the maps from the vacuum solutions in laboratory space onto the media solutions is indispensable. In this context we should mention that these maps, as defined in Refs.~\cite{Leonhardt:2006Nj,Bergamin:2008Pa}, are not unique. As explained in the previous section, generalized transformation optics consists of two steps, a transformation of the equations of motion from laboratory space to the electromagnetic spaces (the diffeomorphisms I and II) and a suitable re-interpretation of the result in laboratory space \footnote{From now on, the notation of generalized transformation optics will be used, unless explicitly mentioned differently. The case of standard transformation optics always follows by a simple identification $\bbar x^\mu = \bar x^\mu$.}. Under the diffeomorphisms the fields transform according to the standard law
\begin{align}
\label{barEB}
 \bar E_i &= \frac{\partial x^0}{\partial \bar x^0} \frac{\partial x^j}{\partial \bar x^i}  E_j\ , & \bar B^i &= \frac{\partial \bar x^i}{\partial x^j} B^j\ , \\
 \label{barDH}
 \bbar D^i &= \bbar{\sigma} \frac{\partial \bbar x^i}{\partial x^j} D_j\ , & \bbar H_i &= \bbar{\sigma} \frac{\partial x^0}{\partial \bbar x^0}\frac{\partial  x^j}{\partial \bbar x^i} H_j\ .
\end{align}
The additional signs $\bbar \sigma$ are needed in order to obtain the correct transformation of the excitation tensor $\mathcal H^{\mu\nu}$ as defined in Eq.\ \eqref{spacevec2} and are a consequence of the factors $\sqrt{-g_{00}}$ in that equation. As coordinate transformations (or local representations of diffeomorphisms) \eqref{barEB} and \eqref{barDH} are unambiguous.

For the second step the rescaling
\begin{align}
\label{scal3}
 \tilde E_i &= \bar s \bar E_i\ , & \tilde B^i &= \bar \sigma \frac{\sqrt{\bar \gamma}}{\sqrt{\gamma}} \bar B^i\ , \\
 \label{scal4}
 \tilde D^i &= \bbar \sigma \frac{\sqrt{-\bbar g}}{\sqrt{-g}} \frac{\partial \bbar x^0}{\partial x^0} \bbar D^i\ , & \tilde H_i &= \bbar s \frac{\sqrt{-g_{00}}}{\sqrt{-\bbar g_{00}}} \frac{\partial \bbar x^0}{\partial x^0} \bbar H_i\ .
\end{align}
has been proposed in Ref.~\cite{Bergamin:2008Pa} as extension of Eqs.~\eqref{scal1} and \eqref{scal2} to generalized transformation optics. Obviously, any change in these equations which does not change the constitutive equations \eqref{tripeleps} and \eqref{tripelmu} represents a physically equivalent though mathematically different identification of the medium solution in terms of vacuum solutions of the Maxwell equations and thus establishes an ambiguity. Two ambiguities are important in the current work:
\begin{itemize}
 \item Diffeomorphisms which leave the metric invariant (so-called isometries) constitute an ambiguity of this kind. They comprise translations, rotations and Lorentz transformations. This ambiguity is important since in the prescription of Refs.~\cite{Leonhardt:2006Nj,Bergamin:2008Pa} it is by no means obvious that a certain point on the interface with coordinates $\tilde x^{\mu}_I$ is represented by the same values of the coordinates $\bar x^{\mu}$ and $\bbar x^{\mu}$ on both sides of the interface. In other words, it is possible that the interface is represented by two different surfaces in the electromagnetic spaces of the two media. This leads to a discontinuity in the vacuum solutions, which however can be removed by a suitable choice of an isometry transformation.
  \item The Maxwell equations are invariant under a rescaling of all fields by a constant factor $\alpha$ and thus this represents an ambiguity in the re-interpretation in laboratory space. We will keep this factor in the following and it will be seen below that it is relevant in the case of negative refractive indices. \footnote{Since the two sets of equations, \eqref{maxwell} and \eqref{maxwell2}, depend on two different sets of fields, a rescaling of the fields of \emph{one} set also represents an invariant transformation of the equations of motion. However, these rescalings change the constitutive relation as they change permittivity and permeability by a (not essentially positive) constant. This implies that the interpretation of a negative refractive index in transformation optics \cite{Leonhardt:2006Nj} actually is the effect of an ambiguity. This is most easily seen in the relativistically covariant formulation: according to  Ref.~\cite{Leonhardt:2006Nj} a negative refractive index is found if $\epsilon^{ijk}$ in Eqs.~\eqref{maxwell} and \eqref{maxwell2} changes sign under the transformation, as this sign has to be absorbed by a rescaling of $\bm E$ and $\bm H$ with a negative constant. However, in the relativistically covariant formulation the (4-dimensional) Levi-Civita symbol only appears as an overall factor in the constraint $\epsilon^{\mu\nu\rho\sigma} \partial_\mu F_{\rho\sigma} = 0$, and thus the change of sign is without any consequences. Thus, starting from the relativistically covariant formulation, the negative refractive index appears rather as an ambiguity.}
\end{itemize}

Combining Eqs.\ \eqref{barEB}, \eqref{barDH} with Eqs.\ \eqref{scal3}, \eqref{scal4} and taking into account the new factor $\alpha$ one obtains
\begin{align}
\label{tildeE}
 \tilde{E}_i\left(\tilde x = \bar x(x)\right) &= \alpha \bar s \frac{\partial x^0}{\partial \bar x^0} \frac{\partial x^j}{\partial \bar x^i} E_j(x)\ , \\
 \label{tildeH}
 \tilde{H}_i\left(\tilde x = \bbar x(x)\right) &= \alpha \bbar s \bbar \sigma \frac{\sqrt{-\bbar g_{00}}}{\sqrt{-g_{00}}} \frac{\partial x^j}{\partial \bbar x^i} H_j(x)\ , \\
 \label{tildeB}
 \tilde B^i\left(\tilde x = \bar x(x)\right) &= \alpha \bar \sigma \frac{\sqrt{\bar \gamma}}{\sqrt{\gamma}} \frac{\partial \bar x^i}{\partial x^j} B^j(x) \ , \\
 \label{tildeD}
 \tilde D^i\left(\tilde x = \bbar x(x)\right) &= \alpha \frac{\sqrt{-\bbar g}}{\sqrt{-g}} \frac{\partial \bbar x^0}{\partial x^0} \frac{\partial \bbar x^i}{\partial x^j} D^j(x) \ .
\end{align}

First we want to show that the transformations of $\bm E$ and $\bm H$ and of $\bm B$ and $\bm D$ are---up to the difference in the associated electromagnetic spaces---equivalent if bi-anisotropic terms in the constitutive relation are absent. Indeed, if $\bar g_{0i} = \bbar g_{0i} = 0$ it easily follows from Eq.~\eqref{dettransform} that
\begin{equation}
\label{g00transform}
 \sqrt{-\bar g_{00}} = \left\lVert \frac{\partial x^0}{\partial \bar x^0} \right\lVert \sqrt{-g_{00}} = \bar \sigma \frac{\partial x^0}{\partial \bar x^0} \sqrt{-g_{00}}
\end{equation}
with a similar relation for $\sqrt{-\bbar{g}_{00}}$ (the symbol $\|a\|$ is used to indicate the absolute value of a number $a$ in order to distinguish it from the determinant of a matrix, $|A|$.) Using this equation with $\sqrt{-g} = \sqrt{-g_{00}}\sqrt{\gamma}$ in Eqs.~\eqref{tildeD} and \eqref{tildeH} straightforwardly establishes the equivalence. Applying the transformation law \eqref{dettransform} to the spatial metric,
\begin{equation}
\label{gammatransform}
 \sqrt{\bar \gamma} =  \sqrt{\left\lvert \frac{\partial x^i}{\partial \bar x^j} \right\lvert^2} \sqrt{\gamma} = \bar s \bar \sigma \left\lvert \frac{\partial x^i}{\partial \bar x^j} \right\lvert \sqrt{\gamma}\ ,
\end{equation}
allows to rewrite Eqs.~\eqref{tildeE}--\eqref{tildeD} as
\begin{align}
\label{tildeEH}
 \tilde{E}_i\left(\tilde x = \bar x(x)\right) &= \alpha \bar s \frac{\partial x^0}{\partial \bar x^0} \frac{\partial x^j}{\partial \bar x^i} E_j(x)\ , \\
 \label{tildeH2}
 \tilde{H}_i\left(\tilde x = \bbar x(x)\right) &= \alpha \bbar s \frac{\partial x^0}{\partial \bbar x^0} \frac{\partial x^j}{\partial \bbar x^i} H_j(x)\ , \\
 \label{tildeBD}
 \tilde B^i\left(\tilde x = \bar x(x)\right) &= \alpha \bar s \left\lvert \frac{\partial x^k}{\partial \bar x^l}\right\lvert \frac{\partial \bar x^i}{\partial x^j} B^j(x) \ , \\
 \label{tildeD2}
 \tilde D^i\left(\tilde x = \bbar x(x)\right) &= \alpha \bbar s \left\lvert \frac{\partial x^k}{\partial \bbar x^l}\right\lvert \frac{\partial \bbar x^i}{\partial x^j} D^j(x) \ .
\end{align}
By means of the relations \eqref{tildeEH}--\eqref{tildeD2} the boundary conditions, which are imposed on the media solutions, can be translated into conditions imposed on the vacuum solutions.

To be able to write down these equations some notations and conventions have to be introduced. Though the interface forms a surface in space, the boundary conditions only will be studied in one particular point of this interface, whose coordinates in laboratory spacetime are denoted by $\tilde x^{i}_I$. Furthermore, it is convenient to choose a certain time instance $\tilde t_I$ as well, since we allow transformations of time. Most of the equations in this section only hold at this specific point in laboratory space, $\tilde x^\mu_I$, which is not indicated specifically if the meaning of the equation is obvious in the context.

\begin{figure}
 \centering
 \includegraphics[width=0.8\linewidth]{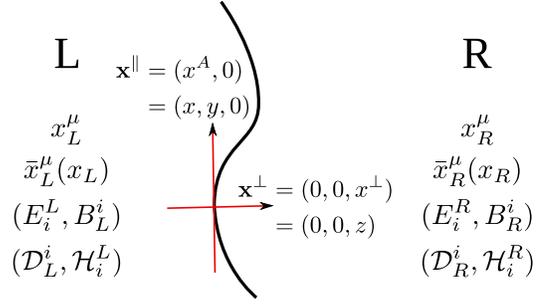}
 % boundary3.eps: 0x0 pixel, 300dpi, 0.00x0.00 cm, bb=0 1 446 300
 \caption{Notations used to describe the interface between two transformation media. The interface between vacuum and a medium follows straightforwardly if one transformation is reduced to the trivial identity map.}
 \label{fig:boundarynot}
\end{figure}

Since most equations are written in an index notation, an adapted coordinate system will be used. Without loss of generality we can assume that at the point $\tilde x^{\mu}_I$ the space vector parallel to the interface is represented as $\bm x_{\parallel} = (x^{A},0)$, where indices with capital Latin letters take values 1,2. The vector perpendicular to the interface accordingly is written as $\bm x_\perp = (0,0,x^{\perp})$. In concrete application we will also use $x^\perp = z$, $(x^A) = (x,y)$. To distinguish the two media we will denote them as \emph{left medium} (index $L$) and \emph{right medium} (index $R$). This situation is also illustrated in Fig.~\ref{fig:boundarynot}.

In our notation the boundary conditions \eqref{BC1} and \eqref{BC2} in absence of surface charges and currents can be written as $\tilde D_L^\perp = \tilde D^\perp_R$, $\tilde B_L^\perp = \tilde B^\perp_R$ and $\tilde E^L_{A} = \tilde E^R_{A}$, $\tilde H^L_{A} = \tilde H^R_{A}$ and with Eqs.~\eqref{tildeEH}--\eqref{tildeD2}
{\allowdisplaybreaks
\begin{gather}
\label{VBCE}
\begin{split}
 \tilde{E}^L_A &= \alpha_L \bar s_L \frac{\partial x^0}{\partial \bar x_L^0} \frac{\partial x^j}{\partial \bar x_L^A} E^L_j=\\ &= \alpha_R \bar s_R \frac{\partial x^0}{\partial \bar x_R^0} \frac{\partial x^j}{\partial \bar x_R^A} E^R_j = \tilde{E}^R_A\ ,
\end{split}\\
\label{VBCH}
\begin{split}
 \tilde{H}^L_A &= \alpha_L \bbar s_L \frac{\partial x^0}{\partial \bbar x_L^0} \frac{\partial x^j}{\partial \bbar x_L^i} H^L_j  =\\& = \alpha_R \bbar s_L \frac{\partial x^0}{\partial \bbar x_R^0} \frac{\partial x^j}{\partial \bbar x_R^i} H^R_j= \tilde{H}^R_A\ , 
\end{split}\\
\label{VBCB}
\begin{split}
  \tilde B^\perp_L &= \alpha_L \bar s_L \left\lvert \frac{\partial x^k}{\partial \bar x_L^l}\right\lvert \frac{\partial \bar x_L^\perp}{\partial x^j} B_L^j =\\&= \alpha_R \bar s_R \left\lvert \frac{\partial x^k}{\partial \bar x_R^l}\right\lvert \frac{\partial \bar x_R^\perp}{\partial x^j} B_R^j = \tilde B^\perp_R\ , 
\end{split}\\
\label{VBCD}
\begin{split}
 \tilde D_L^\perp &= \alpha_L \bbar s_L \left\lvert \frac{\partial x^k}{\partial \bbar x_L^l}\right\lvert \frac{\partial \bbar x_L^\perp}{\partial x^j} D_L^j =\\&= \alpha_R \bbar s_R \left\lvert \frac{\partial x^k}{\partial \bbar x_R^l}\right\lvert \frac{\partial \bbar x_R^\perp}{\partial x^j} D_R^j = \tilde D_R^\perp\ .
\end{split}
\end{gather}
}
These are the boundary conditions in terms of the vacuum solutions which will be considered in the following.

\subsection{Extending the vacuum solution across the interface}
\label{sec:boundaryI}
In this section restrictions on the transformations shall be derived under which the boundary conditions \eqref{BC1} and \eqref{BC2} are satisfied automatically, in other words under which the same vacuum solution $(\bm E, \bm H)$ of the Maxwell equations can be used on both sides of the interface:
\begin{align}
 E^L_i\left(x_L(\tilde x_I)\right) &= E^R_i\left(x_R(\tilde x_I)\right) &\\ H^L_i\left(x_L(\tilde x_I)\right) &= H^R_i\left(x_R(\tilde x_I)\right)
\end{align}
Interfaces of this type disappear on the level of the vacuum solutions and thus they must be reflectionless.

Most of the results of this section have been obtained elsewhere already, in particular Refs. \cite{Yan:2008Pi,Yan:2008Rl}. There are, however, a few differences in the approach taken here: in contrast to Refs.~\cite{Yan:2008Pi,Yan:2008Rl} we intend to construct a solution of the Maxwell equations in the media from the solutions in vacuo, furthermore stretching and eventual inversion of the time direction are included in our calculation and finally we work in the generalized approach of transformation optics according to Ref.\ \cite{Bergamin:2008Pa}.

Since we intend to construct a vacuum solution that extends over the interface, the location of the latter in the electromagnetic spaces must the same for the left and right medium,
\begin{equation}
 \bar x_L^{\mu} (\tilde x_I) = \bar x_R^{\mu} (\tilde x_I)\ , \qquad \bbar x_L^{\mu} (\tilde x_I) = \bbar x_R^{\mu} (\tilde x_I)
\end{equation}
At this point the first ambiguity discussed above is important, since it allows to adjust the two transformations in such a way that this equation holds at the point $\tilde x_I^{\mu}$ without changing the physics. Now we can choose without loss of generality our coordinate system $\tilde x^{\mu}$ according to the discussion above and as illustrated in Fig.~\ref{fig:boundarynot}.

To derive the restrictions on the transformations the map $x^{\mu}\rightarrow\bar{x}^\mu$ is considered first, which affects $\bm E$ and $\bm B$. The discussed extension of the vacuum solution across the interface is required to hold for any solution of the Maxwell equations in vacuum. Therefore, the ensuing restrictions are derived from \eqref{VBCE} and \eqref{VBCB} as
\begin{align}
\label{parallelI}
 \alpha_L \bar s_L \frac{\partial x^0}{\partial \bar x_L^0} \frac{\partial x^j}{\partial \bar x_L^A} &= \alpha_R \bar s_R \frac{\partial x^0}{\partial \bar x_R^0} \frac{\partial x^j}{\partial \bar x_R^A}\ , \\
 \label{normalI}
 \alpha_L \bar s_L \left\lvert \frac{\partial x^k}{\partial \bar x_L^l}\right\lvert \frac{\partial \bar x_L^\perp}{\partial x^j} &= \alpha_R \bar s_R \left\lvert \frac{\partial x^k}{\partial \bar x_R^l}\right\lvert \frac{\partial \bar x_R^\perp}{\partial x^j}\ .
\end{align}
By multiplying Eq.\ \eqref{parallelI} by $\partial \bar x_L^B/\partial x^j$ and summing over $j$ this equation yields the conditions
\begin{align}
\label{gsI}
 \frac{\partial \bar x_L}{\partial \bar x_R} &= \frac{\partial \bar y_L}{\partial \bar y_R} = \bar s_L \bar s_L \frac{\alpha_L}{\alpha_R} \frac{\partial \bar x^0_R}{\partial \bar x^0_L}\ , & \frac{\partial \bar x_L}{\partial \bar y_R} &= \frac{\partial \bar y_L}{\partial \bar x_R} = 0\ .
\end{align}
As the maps shall be continuous, the two transformations must agree along the boundary and thus
\begin{equation}
 \bar s_L \bar s_R \frac{\alpha_L}{\alpha_R} \frac{\partial \bar x^0_R}{\partial \bar x^0_L} = 1\ .
\end{equation}
To simplify condition \eqref{normalI} the relation $\lvert\partial x^i/\partial \bar x_R^j\lvert/\lvert\partial x^k/\partial \bar x_L^l\lvert = \lvert\partial \bar x_L^i/\partial \bar x_R^j\lvert$ may be used. Then the restriction on the transformation of the normal component $\bar x^{\perp}$ can be written as
\begin{equation}
 \frac{\partial \bar x_L^\perp}{\partial \bar x_R^\perp} = \bar s_L \bar s_L \frac{\alpha_R}{\alpha_L}\left\lvert \frac{\partial \bar x_L^k}{\partial \bar x_R^l}\right\lvert = \frac{\partial \bar x^0_R}{\partial \bar x^0_L} \left\lvert \frac{\partial \bar x_L^k}{\partial \bar x_R^l}\right\lvert\ .
\end{equation}
By virtue of Eq.~\eqref{gsI} the determinant reduces to
\begin{equation}
\label{gsdetconstr}
 \left\lvert \frac{\partial \bar x_L^k}{\partial \bar x_R^l}\right\lvert = \frac{\partial \bar x^\perp_L}{\partial \bar x^\perp_R} - \frac{\partial \bar x^\perp_L}{\partial \bar x^A_R} \frac{\partial \bar x^A_L}{\partial \bar x^\perp_R} = \frac{\partial \bar x^\perp_L}{\partial \bar x^\perp_R}\ ,
\end{equation}
where the fact that $\partial x^i/\partial \bar x^A_L = \partial x^i/\partial \bar x^A_R$ has been used. The different restrictions \eqref{gsI}--\eqref{gsdetconstr} now can be summarized in the following simple form:
\begin{align}
\label{simple}
 \frac{\partial \bar x^A_L}{\partial \bar x^B_R} &= \delta^A_B & \frac{\partial \bar x^0_L}{\partial \bar x^0_R} &= 1
\end{align}
$\partial \bar x_L^\perp/\partial \bar x_R^\perp$ remains unrestricted. There exist no transformations with stretchings and/or reversal of time that allow an extension over an interface. Still, space inversions of the type $\bar x^\perp_L = - \beta x^\perp_R$ are possible and in these cases $\alpha_R/\alpha_L = -1$. Without loss of generality it then can be assumed that $\alpha = \pm 1$ in all mappings.

The whole calculation needs to be redone for the fields $\bm D$ and $\bm H$. This generates a new set of conditions, which is found simply by replacing all barred variables in Eqs.~\eqref{parallelI}--\eqref{simple} by double-barred ones. An important comment is in order: since the global factors $\alpha_R$ and $\alpha_L$ only can be chosen once, they need to be the same for both mappings. Thus solutions only extend over an interface if either none or both mappings include space inversions. Thus, it is seen that indeed the ambiguity associated with the constant $\alpha$ plays an important role in the discussion of boundary conditions as soon as negative refractive index media are involved.

Although many proposals of transformation designed devices, most importantly the invisibility cloak \cite{Pendry:2006Sc,Leonhardt:2006Sc} and the perfect lens \cite{Leonhardt:2006Nj}, fit into the picture described in this section, it is important to realize that in many interesting situations this setup might be too limited. In many situations transformation designed devices do not exclusively contain reflectionless boundaries. This can happen if some surfaces do not contribute to the functionality of the device. In this case this might be regarded as a minor problem since the above calculation still applies locally. However, one also might deal with reflections at functional surfaces, which requires an extension of the approach.

\subsection{General transformations}
\label{sec:boundaryII}
In this section interfaces between two arbitrary generalized transformation media are considered. The transformations at such interfaces do not necessarily obey the restrictions \eqref{parallelI} and \eqref{normalI} and consequently it can no longer be required that a certain solution of the Maxwell equations in the two media is described by the same vacuum solution. Of course, a specific solution for $\bm E$ and $\bm H$ via Eqs.~\eqref{tildeEH}--\eqref{tildeD2} still provides solutions of the Maxwell equations on both sides of the interface, but we no longer insist that these solutions obtained from the \textit{same} vacuum solution meet the boundary conditions \eqref{BC1} and \eqref{BC2}. This allows to relax all constraints found in the previous section, in particular the transformations need not even be continuous at the interface. It will be shown in the following how the boundary conditions \eqref{BC1} and \eqref{BC2} (still with $\sigma = 0$ and $\bm K = 0$) can be rewritten in terms of the vacuum solutions in laboratory space and the geometric transformations.

Let us start with the result already obtained in Eqs.~\eqref{VBCE}--\eqref{VBCD}. In this section these equations are no longer seen as restrictions onto the transformations, but rather define the free space solution $(E_i^L,H_i^L)$ at the interface in terms of $(E_i^R,H_i^R)$ taken at this point. A complication arises as Eqs.~\eqref{VBCE} and \eqref{VBCH} relate field components with lower indices in laboratory space, while Eqs.~\eqref{VBCB} and \eqref{VBCD} relate upper indices. Due to Eqs.~\eqref{scal3} and \eqref{scal4} this also applies in electromagnetic space, where the boundary conditions can be reformulated as
\begin{gather}
\label{nonGS1}
\tilde E^L_A = \tilde E^R_A \quad \Rightarrow \quad \bar{E}^L_A = \bar s_L \bar s_R \frac{\alpha_R}{\alpha_L} \bar{E}^R_A \ ,\\
\label{nonGS2}
 \tilde B^\perp_L = \tilde B^\perp_R \quad \Rightarrow \quad \bar B^\perp_L  = \bar s_L \bar s_R \frac{\alpha_R}{\alpha_L} \left\lvert \frac{\partial \bar x_{L}^k}{\partial \bar x^l_R}\right\lvert \bar B^\perp_R \ , \\
 \label{nonGS3}
 \tilde{H}^L_A = \tilde{H}^R_A \quad \Rightarrow \quad \bbar{H}^L_A = \bbar s_L \bbar\sigma_L \bbar s_R \bbar\sigma_R \frac{\alpha_R}{\alpha_L}  \bbar{H}^R_A \ ,\\
 \label{nonGS4}
 \tilde D^\perp_L = \tilde D^\perp_R \quad \Rightarrow \quad \bbar D^\perp_L  = \bbar s_L \bbar \sigma_L \bbar s_R\bbar\sigma_R \frac{\alpha_R}{\alpha_L} \left\lvert \frac{\partial \bbar x_{L}^k}{\partial \bbar x^l_R}\right\lvert \bbar D^\perp_R \ .
\end{gather}
Here, all fields are taken at those points which are mapped onto the interface in laboratory space, e.g., $\bar E_A^L$ and $\bar B^\perp_L$ are taken at the spacetime point $\left(\bar t_L(\tilde t_I), \bar x^i_{L}(\tilde x_I^j)\right)$. As we no longer insist on continuous mappings this spacetime point in laboratory space may be represented by two different spacetime points in electromagnetic space on the two sides of the interface. Since no a priori assumptions about the metric in the electromagnetic spaces should be made, the derivation of the boundary conditions exclusively in upper (or alternatively lower) indices is not straightforward. To simplify this task standard transformation optics \cite{Leonhardt:2006Nj,Leonhardt:2008Oe} is considered in a first step.

\subsubsection{Standard transformation optics}
In this subsection the boundary conditions are discussed for standard transformation optics and thus $\bar x^\mu \equiv \bbar x^\mu$ holds. Then with Eqs.~\eqref{parallelmetric} and \eqref{parallelmetricII} the following relations can be established:
\begin{align}
\label{STO1}
 \bar E^A &= \left(\bar g^{AB} - \frac{\bar g^{A\perp} \bar g^{\perp B}}{\bar g^{\perp\perp}}\right) \bar E_A + \frac{\bar g^{A\perp}}{\bar g^{\perp\perp}} \bar E^\perp \\ \label{STO1.1}
 \bar D_\perp &= \frac{1}{\bar g^{\perp \perp}}(\bar D^\perp - \bar g^{\perp A} \bar D_A)
\end{align}
These two relations enable us to rewrite the boundary conditions in electromagnetic space (Eqs.~\eqref{nonGS1}--\eqref{nonGS4}) exclusively in terms of vectors (lower indices) or covectors (upper indices). If we intend to express everything in terms of vectors one uses the relation \eqref{STO1.1} and after some algebra arrives at
\begin{align}
\label{STO2}
 E_i^L &= \bar s_L \bar s_R \frac{\alpha_R}{\alpha_L} \frac{\partial \bar x^0_L}{\partial \bar x^0_R} V_{i}{}^{k} \frac{\partial x_R^j}{\partial \bar x^k_R} E_j^R\ ,\\
 \label{STO3}
 H_i^L &= \bar s_L \bar s_R \frac{\alpha_R}{\alpha_L} \frac{\partial \bar x^0_L}{\partial \bar x^0_R}  V_{i}{}^{k} \frac{\partial x_R^j}{\partial \bar x^k_R} H_j^R\ ,
\end{align}
where $V_{i}{}^{k}$ is given by
\begin{equation}
 V_i{}^{k} = \frac{\partial \bar x^k_L}{\partial x_L^i} + \frac{1}{\bar g^{\perp\perp}_L} \frac{\partial \bar x^\perp_L}{\partial x^i_L}\left(\left\lvert\frac{\partial \bar x^m_L}{\partial \bar x^n_R}\right\lvert \frac{\partial \bar x^0_R}{\partial \bar x^0_L} \bar g^{\perp k}_R - \bar g^{\perp k}_L \right)\ .
\end{equation}
For covectors raising of all indices implies
\begin{align}
\label{STO4}
 E^i_L &= \bar s_L \bar s_R \frac{\alpha_R}{\alpha_L} \frac{\partial x^i_L}{\partial \bar x^k_L} C^{k}{}_{j} E^j_R\ , \\
  \label{STO5}
  H^i_L &= \bar s_L \bar s_R \frac{\alpha_R}{\alpha_L} \frac{\partial x^i_L}{\partial \bar x^k_L} C^{k}{}_{j} H^j_R\ ,
\end{align}
with
\begin{equation}
\label{Ckj}
\begin{split}
 C^{k}{}_{j} &=\left(\bar g_L^{kA} - \frac{\bar g_L^{k\perp}\bar g_L^{\perp A}}{\bar g_L^{\perp\perp}}\right) \bar g^R_{Al}\frac{\partial \bar x_R^l}{\partial x^j_R}+\\ &\quad + \left\lvert\frac{\partial \bar x^m_L}{\partial \bar x^n_R}\right\lvert \frac{\partial \bar x^0_R}{\partial \bar x^0_L} \frac{\bar g_L^{k\perp}}{\bar g^{\perp\perp}_L} \frac{\partial \bar x_R^\perp}{\partial x^j_R}\ .
\end{split}
\end{equation}
In these equations it is important to remember that $\bar E^i = \sqrt{-\bar g_{00}} \bar D^i$ and $\bar H^i = \sqrt{-\bar g_{00}} \bar B^i$. Furthermore, notice that the fields on both sides of these equations are taken at the same value in laboratory space, $\tilde x_I^\mu$, cf.\ Eqs.\ \eqref{nonGS1}--\eqref{nonGS4}. Eqs.~\eqref{STO2}--\eqref{Ckj} are the reformulation of the boundary conditions in terms of the vacuum solutions in laboratory space. As can be seen, the field values at the interface of the vacuum solution of the left medium are completely defined in terms of the vacuum solution of the right medium at this point and the geometric manipulations of transformation optics. The equations do no longer include any reference to the medium solutions, which actually describe the physical situation.

\paragraph{Example: homogeneous and isotropic media}

To show how basic characteristics of media are encoded in Eqs.~\eqref{STO2}--\eqref{Ckj} let us consider a simple example, the flat interface between vacuum and a homogeneous and isotropic medium. Media of this type with $\epsilon = \mu = n$ can be obtained by the simple transformation $\bar t = n t$, as is seen when inserting this relation into Eqs.~\eqref{epstilde2} and \eqref{mutilde2}. Unless $n = \pm 1$ the interface is not reflectionless, however the above conditions still describe correctly the boundary conditions that have to hold.

\begin{figure}
 \centering
 \includegraphics[width=0.8\linewidth,bb=0 0 253 154]{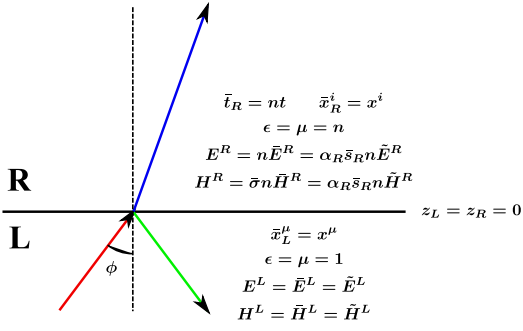}
 % hom_isotr_medium.eps: 149740240x-1243926160 pixel, 300dpi, 1267800.75x-10531908.00 cm, bb=0 0 253 154
 \caption{Interface between vacuum and a homogeneous and isotropic medium according to transformation optics.}
 \label{fig:HomIsoMed}
\end{figure}
For concreteness let us assume the situation as depicted in Fig.\ \ref{fig:HomIsoMed}. Since the value of $\alpha_R$ can be assumed to be $\pm 1$ we will choose it in such a way that $\bar s_R \alpha_R = 1$. With this choice Eqs.~\eqref{STO2} and \eqref{STO3} reduce to
\begin{align}
\label{STO6}
 E_A^L &= \frac{1}{n} E_A^R\ , & E_\perp^L &=  E_\perp^R\ , \\
 \label{STO7}
 H_A^L &= \frac{1}{n} H_A^R\ , & H_\perp^L &= H_\perp^R\ .
\end{align}
To study the reflection and refraction coefficients we start by defining the solution in the right medium. Its vacuum solution is assumed as a plane wave,
\begin{align}
\label{STO7.1}
 \bm E^R &= \bm e \exp\left[i(\bm k_R \cdot \bm x_R - \omega_R t_R)\right] + \mbox{c.c.}\ ,\\ \bm H^R &= \bm h \exp\left[i(\bm k_R \cdot \bm x_R - \omega_R t_R)\right] + \mbox{c.c.}\ ,
\end{align}
which is mapped onto the solution in the medium as $\tilde{\bm E}^R = \bm E^R/n$, $\tilde{\bm H}^R = \bm H^R/n$. In the simple example given here it is immediate that the plane wave of the vacuum solution obeying $\bm k_R^2 = \omega_R^2$ maps onto a plane wave in the medium with dispersion relation $\tilde{\bm k}_R^2 = n^2 \tilde{\omega}_R^2$, as required by the fact that $\epsilon = \mu = n$.

From Eqs.~\eqref{STO6} and \eqref{STO7} the solution for $\bm E^L$ and $\bm H^L$ at the interface is deduced as
\begin{align}
\begin{split}
 E_A^L &= \frac{1}{n} e_A \exp\left[i(\bm k_R \cdot \bm x_R - \omega_R t_R)\right]\\ & = \frac{1}{n} e_A \exp\left[i(\bm k_L \cdot \bm x_L - \frac{1}{n}\omega_L t_L)\right]\ ,
 \end{split}
  \\
\begin{split}
 E_\perp^L &= e_\perp \exp\left[i(\bm k_R \cdot \bm x_R - \omega_R t_R)\right] \\ &= e_\perp   \exp\left[i(\bm k_L \cdot \bm x_L - \frac{1}{n} \omega_L t_L)\right]\ ,
 \end{split} \\
\begin{split}
 H_A^L &= \frac{1}{n} h_A \exp\left[i(\bm k_R \cdot \bm x_R - \omega_R t_R)\right] \\&= \frac{1}{n} h_A \exp\left[i(\bm k_L \cdot \bm x_L - \frac{1}{n} \omega_L t_L)\right]\ ,
 \end{split} \\
\begin{split}
 H_\perp^L &=  h_\perp \exp\left[i(\bm k_R \cdot \bm x_R - \omega_R t_R)\right] \\ &= h_\perp \exp\left[i(\bm k_L \cdot \bm x_L - \frac{1}{n} \omega_L t_L)\right]\ .
 \end{split}
\end{align}
While the solution $(\bm e, \bm h, \bm k_R)$ by construction obeys $\bm k_R \cdot \bm e = \bm k_R \cdot \bm h = \bm e \cdot \bm h = 0$ and in addition $\bm k_R^2 = \omega_R^2$, the corresponding relations for the solution in the vacuum (``left hand side'') are not immediate. It is easily seen that they cannot be met simultaneously with a single plane wave solution unless $e_\perp = h_\perp = k^R_A = 0$, in other words unless the incoming wave hits the interface at normal incidence.

For simplicity let us assume in the following that the magnetic field is perpendicular to the plane of incidence, i.e.\ $h_\perp = 0$. Then $\bm H^L$ automatically is perpendicular to the wave vector, while for the electric field we make the ansatz
\begin{align}
\label{STO10}
 E_A^L &= \left(S+(1-S)\right) E_A^L = (E_{\mbox{\tiny in}})_A + (E_{\mbox{\tiny ref}})_A\ , \\
 \label{STO10.1} E_\perp^L &= \left(T+(1-T)\right) E_\perp^L = (E_{\mbox{\tiny in}})_\perp + (E_{\mbox{\tiny ref}})_\perp\ ,
\end{align}
where $S$ and $T$ are chosen in such a way that $\bm k_{\mbox{\tiny in}} \cdot \bm e_{\mbox{\tiny in}} = 0$ and $\bm k_{\mbox{\tiny ref}} \cdot \bm e_{\mbox{\tiny ref}} = 0$. As indicated by the notation in Eqs.~\eqref{STO10} and \eqref{STO10.1}, the first solution represents the incoming wave while the second one is the reflected one. If we choose $\tilde z = z_R = z_L = 0$ as location of the interface this implies
\begin{align}
S \frac{e_A}{n} k^L_A + T e_\perp k^L_\perp &= 0\ , \\ (1-S) \frac{e_A}{n} k^L_A -(1-T) e_\perp k^L_\perp &= 0\ ,\\ \label{kldisprel} \bm k_L^2 = \omega_L^2 &= \frac{\omega_R^2}{n^2}\ .
\end{align}
Eq.\ \eqref{kldisprel} together with $\bm k_R^2 = \omega_R^2$ allows to deduce
\begin{equation}
(k_\perp^L)^2 = (k_\perp^R)^2 + (\frac{1}{n^2}-1) \omega_R^2\ .
\end{equation}
With this one finds for $S$ and $T$
\begin{align}
 S &= \frac{1}{2}\frac{n \cos \phi + \sqrt{n^2-\sin^2\phi}}{\sqrt{n^2-\sin^2\phi}}\ , \\ T&= \frac{1}{2} \frac{n \cos \phi + \sqrt{n^2-\sin^2\phi}}{n\cos\phi}\ .
\end{align}
From these equations and the dispersion relation of $\bm k_L$ it is now easy to show that
\begin{align}
\label{STO8}
 \frac{\|\bm E_{\mbox{\tiny trans}}\|}{\|\bm E_{\mbox{\tiny in}}\|} &= \frac{2n^2\cos \phi}{n \cos\phi + \sqrt{n^2-\sin^2\phi}}\ , \\
 \label{STO8.1}
\frac{\|\bm E_{\mbox{\tiny ref}}\|}{\|\bm E_{\mbox{\tiny in}}\|} &= \frac{n\cos \phi - \sqrt{n^2-\sin^2\phi}}{n \cos\phi + \sqrt{n^2-\sin^2\phi}} \ ,
\end{align}
where $\bm E_{\mbox{\tiny trans}}$ is the solution \eqref{STO7.1}. Notice that these relations determine the reflection and transmission coefficients in terms of the vacuum solutions of transformation optics. From the relation $\tilde{\bm E}^R = \bm E^R/n$ it can be seen that the additional factor $n$ in the first relation indeed reproduces the correct result in terms of the solutions in laboratory space.

The polarization with $e_{\bot} = 0$ follows analogously by taking the equations for $\bm H$ instead of those for $\bm E$. Obviously this yields again \eqref{STO8} and \eqref{STO8.1} as is required since $\epsilon = \mu$ in our example.

\subsubsection{Generalized transformations}

The situation gets slightly more complicated in generalized transformation optics \cite{Bergamin:2008Pa}, i.e.\ if the constraint $\bar x^{\mu} \equiv \bbar x^{\mu}$ is relaxed. The complete boundary conditions in electromagnetic space are deduced for the electric field from Eqs.~\eqref{nonGS1} and \eqref{nonGS4}, those for the magnetic field from Eqs.~\eqref{nonGS2} and \eqref{nonGS3}. As is seen, the boundary condition for the lower parallel components are formulated in a different spacetime than the ones for the upper normal component. To reformulate all boundary conditions in terms of the same electromagnetic space, the condition \eqref{nonGS4} with help of the transformation \eqref{barDH} can be rewritten as
\begin{equation}
\label{Dbbarbar}
 \frac{\partial \bbar{x}^{\perp}_L}{\partial \bar x^i_L} \bar \gamma^{ij}_L \bar D_j^L  = \bbar s_L \bar \sigma_L \bbar s_R\bar\sigma_R \frac{\alpha_R}{\alpha_L} \left\lvert \frac{\partial \bbar x_{L}^k}{\partial \bbar x^l_R}\right\lvert \frac{\partial \bbar{x}^{\perp}_R}{\partial \bar x^i_R} \bar \gamma^{ij}_R \bar D_j^R\ .
\end{equation}
This additional transformation between the two electromagnetic spaces induces a mixing of the two transformations in the ensuing boundary condition. Now, Eqs.~\eqref{nonGS1} and \eqref{Dbbarbar} can be combined to derive the boundary conditions in terms of the vacuum solutions in laboratory space in analogy to Eq.~\eqref{STO2}. Using the notation \eqref{gbbb} it can be cast into the rather simple form
\begin{equation}
\label{TSM1}
 E_i^L = \frac{\alpha_R}{\alpha_L} \frac{\partial \bar x^0_L}{\partial \bar x^0_R}  \bar V_{i}{}^{k} \frac{\partial x_R^j}{\partial \bar x^k_R} E_j^R\ ,
\end{equation}
where the new transformation matrix $\bar V_{i}{}^{k}$ takes the form
\begin{multline}
\label{Vbar}
 \bar V_{i}{}^{k} = \bar s_L\bar s_R \frac{\partial \bar x^k_L}{\partial x_L^i} + \\ + \frac{1}{g^{\bbar \perp \bar \perp}_L} \frac{\partial \bar x^\perp_L}{\partial x^i_L}\left(\bbar s_L \bbar s_R \left\lvert\frac{\partial \bbar x^m_L}{\partial \bbar x^n_R}\right\lvert \frac{\partial \bar x^0_R}{\partial \bar x^0_L} g^{\bbar \perp \bar k}_R - \bar s_L \bar s_R  g^{\bbar \perp \bar k}_L \right)\ .
\end{multline}
The calculation of the boundary conditions of $\bm H$ follows analogously by interchanging $\bar x^\mu \leftrightarrow \bbar x^\mu$:
\begin{equation}
\label{TSM2}
 H_i^L = \frac{\alpha_R}{\alpha_L} \frac{\partial \bbar x^0_L}{\partial \bbar x^0_R} \bbar V_{i}{}^{k}  \frac{\partial x_R^j}{\partial \bbar x^k_R} H_j^R\ .
\end{equation}
Here, $\bbar V_{i}{}^{k}$ is obtained from the expression \eqref{Vbar} by replacing all barred quantities and indices by double-barred and vice versa.

\paragraph{More on homogeneous and isotropic media} In the previous section homogeneous and isotropic media from standard transformation optics were considered, which follow from time stretchings $\bar t = n t$ and describe media with $\epsilon  = \mu = n$. This suggests to consider the transformations
\begin{align}
\label{TSM4}
 \bar t &= \mbox{sgn}(\mu) \|\epsilon\| t\ , & \bbar t &= \mbox{sgn}(\epsilon) \|\mu\| t
\end{align}
within the generalized setup, where $\mbox{sgn}(a)$ is the sign of $a$. Applying this transformation in \eqref{tripeleps} and \eqref{tripelmu} yields
\begin{align}
\label{TSM5}
 \tilde D^i &= \epsilon \gamma^{ij} \tilde E_j\ , & \tilde B^i &= \mu \gamma^{ij} \tilde H_j\ ,
\end{align}
which explains the choice of signs in Eq.\ \eqref{TSM4}. This shows that homogeneous and isotropic media with arbitrary permittivity and permeability can be understood as an independent stretching of time in the two different transformations. On a side-remark we notice that these media can also be obtained by stretching all spatial directions simultaneously. Indeed, the transformation
\begin{align}
 \bar x^i &= \frac{\mbox{sgn}(\epsilon) \mbox{sgn}(\mu)}{\|\epsilon\|^{\frac{1}{3}} \|\mu\|^{\frac{2}{3}}} x^i\ , & \bbar x^i &= \frac{\mbox{sgn}(\epsilon) \mbox{sgn}(\mu)}{\|\epsilon\|^{\frac{2}{3}} \|\mu\|^{\frac{1}{3}}} x^i
\end{align}
also yields the media properties \eqref{TSM5}.

\begin{figure}
 \centering
 \includegraphics[width=0.8\linewidth,bb=0 0 253 154]{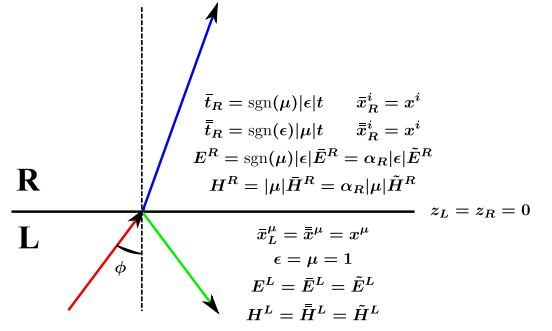}
 % hom_isotr_medium.eps: 149740240x-1243926160 pixel, 300dpi, 1267800.75x-10531908.00 cm, bb=0 0 253 154
 \caption{Interface between vacuum and a homogeneous and isotropic medium with arbitrary and independent $\epsilon$ and $\mu$ according to generalized transformation optics.}
 \label{fig:HomIsoMedII}
\end{figure}
Extending the result of the example of the previous section, we want to derive the law of reflection and refraction at an interface between vacuum and an arbitrary homogeneous, isotropic medium. This situation is depicted in Fig.~\ref{fig:HomIsoMedII}. Again, the calculation starts from the definition of the vacuum solution that is mapped onto the transmitted wave
\begin{align}
 \bm E^R &= \bm e \exp\left[i(\bm k_R \cdot \bm x_R - \omega_R t_R)\right] + \mbox{c.c.}\ ,\\ \bm H^R &= \bm h \exp\left[i(\bm k_R \cdot \bm x_R - \omega_R t_R)\right] + \mbox{c.c.}\ .
\end{align}
Before considering the boundary conditions it might be useful to derive in detail how this vacuum solution is mapped onto a solution of the medium. In a first step we apply the transformations \eqref{TSM4}, which maps $\bm E$ and $\bm H$ onto the solutions (notice the relation $\mbox{sgn}(\mu) = \bar s$, $\mbox{sgn}(\epsilon) = \bbar s$)
\begin{align}
 \bar{\bm E}^R(\bar x, \bar t) &= \frac{\bar s}{\|\epsilon\|}\bm e \exp\left[i(\bar{\bm k}_R \cdot \bar{\bm x}_R - \bar{\omega}_R \bar{t}_R)\right] + \mbox{c.c.}\ , \\
 \bbar{\bm H}^R(\bbar x, \bbar t) &= \frac{1}{\|\mu\|}\bm h \exp\left[i(\bbar{\bm k}_R \cdot \bbar{\bm x}_R - \bbar{\omega}_R \bbar{t}_R)\right] + \mbox{c.c.}\ ,
\end{align}
with $\bar{\bm k}_R^2 =  \epsilon^2 \bar \omega_R^2$ and $\bbar{\bm k}_R^2 =  \mu^2 \bbar \omega_R^2$.
These solutions are now re-interpreted in terms of laboratory space,
\begin{align}
\label{TSM6}
 \tilde{\bm E}^R(\tilde x, \tilde t) &= \frac{\alpha_R}{\|\epsilon\|}\bm e \exp\left[i(\tilde{\bm k}_R \cdot \tilde{\bm x}_R - \tilde{\omega}_R \tilde{t}_R)\right] + \mbox{c.c.}\ ,\\
 \label{TSM7}
 \tilde{\bm H}^R(\tilde x, \tilde t) &= \frac{\alpha_R}{\|\mu\|}\bm h \exp\left[i(\tilde{\bm k}_R \cdot \tilde{\bm x}_R - \tilde{\omega}_R \tilde{t}_R)\right] + \mbox{c.c.}\ ,
\end{align}
with $\tilde{\bm k}_R^2 =  \epsilon \mu \tilde \omega_R^2$. Although the correct dispersion relation is not present in the solutions in electromagnetic space, it is important to notice that also this information is encoded in a completely geometric way, since the general dispersion relation (in the absence of bi-anisotropic contributions to the constitutive relation) reads \cite{Bergamin:2008Pa}: $g^{\bar i \bbar j} k_i k_j = - g^{\bar 0 \bbar 0} \omega^2$.

The calculation of reflection and refraction coefficients follows in close analogy to the example presented above. As boundary conditions we find from \eqref{TSM1} and \eqref{TSM2}
\begin{align}
\label{TSM10}
 E_A^L &= \frac{\alpha_R \bar s_R}{\mbox{sgn}(\mu) \|\epsilon\|} E_A^R\ , & E_\perp^L &=  \alpha_R \bbar s_R E_\perp^R\ , \\
 \label{TSM11}
 H_A^L &= \frac{\alpha_R \bbar s_R}{\mbox{sgn}(\epsilon)\|\mu\|} H_A^R\ , & H_\perp^L &= \alpha_R \bar s_R H_\perp^R\ .
\end{align}
A possible simple choice of $\alpha_R$ is $\alpha_R = \bbar s_R$ which implies
\begin{align}
 E_A^L &= \frac{1}{\epsilon} e_A \exp\left[i(\bm k_L \cdot \bm x_L - \frac{1}{n}\omega_L t_L)\right]\ , \\
 E_\perp^L &=  e_\perp \exp\left[i(\bm k_L \cdot \bm x_L - \frac{1}{n} \omega_L t_L)\right]\ , \\
 H_A^L &=  \frac{\bar s \bbar s}{\mu} h_A \exp\left[i(\bm k_L \cdot \bm x_L - \frac{1}{n} \omega_L t_L)\right]\ , \\
 H_\perp^L &=  \bar s \bbar s h_\perp \exp\left[i(\bm k_L \cdot \bm x_L - \frac{1}{n} \omega_L t_L)\right]\ ,
\end{align}
since $\bm k_R = \bm k_L$ and $\omega_R = \omega_L/n$ with $n = \sqrt{\epsilon \mu}$. If $h_\perp = 0$ the ansatz \eqref{STO10} and \eqref{STO10.1} yields as conditions for $S$ and $T$
\begin{align}
S \frac{e_A}{\epsilon} k^L_A + T e_\perp k^L_\perp &= 0\ , \\ (1-S) \frac{e_A}{\epsilon} k^L_A -(1-T) e_\perp k^L_\perp &= 0\ , \\ \bm k_L^2 = \omega_L^2 &= \frac{\omega_R^2}{n^2}\ .
\end{align}
From these conditions it is found that
\begin{align}
 S &= \frac{1}{2}\frac{\epsilon \cos \phi + \sqrt{n^2-\sin^2\phi}}{\sqrt{n^2-\sin^2\phi}}\ , \\ T&= \frac{1}{2} \frac{\epsilon \cos \phi + \sqrt{n^2-\sin^2\phi}}{\epsilon \cos\phi}\ ,
\end{align}
and thus the generalization of the result \eqref{STO8}, \eqref{STO8.1} is easily derived as
\begin{align}
\label{TSM12}
 \frac{\|\bm E_{\mbox{\tiny trans}}\|}{\|\bm E_{\mbox{\tiny in}}\|} &= \frac{2\epsilon n \cos \phi}{\epsilon \cos\phi + \sqrt{n^2-\sin^2\phi}}\ , \\
 \label{TSM13}
 \frac{\|\bm E_{\mbox{\tiny ref}}\|}{\|\bm E_{\mbox{\tiny in}}\|} &= \frac{\epsilon \cos \phi - \sqrt{n^2-\sin^2\phi}}{\epsilon \cos\phi + \sqrt{n^2-\sin^2\phi}} \ .
\end{align}
This is the law of reflection and refraction in terms of vacuum solutions. Keeping in mind that from Eq.~\eqref{TSM6}, with our choice of $\alpha_R$, $\tilde{\bm E}^L = \bm E^L/\epsilon$ it is seen that this result indeed reproduces the correct law in terms of the laboratory space solutions. Again, the case $e_\bot = 0$ follows completely analogously.

In this example it has been shown that homogeneous and isotropic media with arbitrary permittivity and permeability allow a geometric interpretation in terms of generalized transformation optics. This does not just include the derivation of solutions in these media from vacuum solutions, but also the laws of reflection and refraction at an interface.

\section{Concluding remarks}
\label{sec:conclusions}

In this paper boundary conditions at the interface of two generalized transformation media have been studied and it has been shown how the ensuing conditions can be expressed completely in terms of vacuum solutions of the Maxwell equations and of geometric manipulations. This task has been carried out in two steps: in a first step we considered the most general situation that allows to describe the electromagnetic fields on both sides of the interface in terms of the same vacuum solution. Obviously, all interfaces of this type are reflectionless. In a second step we relaxed the condition of a single vacuum solution and considered interfaces between two arbitrary transformation media. Of course, these are not reflectionless in general; as basic example we showed how generalized transformation optics allows to derive the standard law of reflection and refraction at the interface of homogeneous and isotropic media in a geometric way.

Once this second step has been calculated it is worth to reconsider the meaning of the first result. We have shown that the boundary conditions at a generic (not necessarily reflectionless) interface can be described completely in terms of vacuum solutions of the Maxwell equations in laboratory space and geometric manipulations (diffeomorphisms, locally interpreted as coordinate transformations.) It is well known that diffeomorphisms have a group structure and it is evident that this group structure extends straightforwardly to transformation media: two transformations applied subsequently again yield a transformation that describes a transformation medium, there exist a unit element, which is the trivial transformation, and to each transformation there exists an inverse which reverts the action of the former.

Given an interface between two generalized transformation media, whose boundary conditions in terms of vacuum solutions are described by Eqs.~\eqref{TSM1} and \eqref{TSM2}, we may ask the following question: which interfaces between two different transformation media yield the \emph{same} boundary conditions in terms of the vacuum solutions? The answer is given by the result of Sect.~\ref{sec:boundaryI}. From our specific choice of media we can obtain physically different situations by applying to these media solutions additional transformations that obey the constraints \eqref{parallelI} and \eqref{normalI}. Under such transformations the Eqs.~\eqref{TSM1} and \eqref{TSM2} do not change and thus the boundary conditions in terms of the vacuum solutions are not changed. Therefore the transformations of Sect.~\ref{sec:boundaryI} define equivalent classes of transformation media within the general result of Sect.~\ref{sec:boundaryII} and with respect to the group structure of diffeomorphisms. The reflectionless media are those which are members of the equivalent class of an interface vacuum-vacuum.

Finally it should be mentioned that similar formulas could also be derived for bi-anisotropic media. Still, since the coordinate transformations associated with these media mix the spatial directions with the time direction in order to obtain $\bar g_{0i} \neq 0$ and $\bbar g_{0i} \neq 0$, they will also mix $\bm E$ with $\bm B$ and $\bm D$ with $\bm H$. Thus the ensuing conditions are expected to be considerably more involved.

\begin{acknowledgments}
The author wishes to thank J.~Llorens Montolio for helpful discussions. This work profited a lot from fruitful discussion with M.~Qiu and M.~Yan and W.~Yan during a cooperation of the Advanced Concepts Team of the European Space Agency with the Royal Institute of Technology (KTH). The cooperation was funded under the Ariadna program of ESA.
\end{acknowledgments}

\appendix*
\section{Covariant formulation}
\label{sec:conventions}
In this Appendix we present our notations and conventions regarding the covariant formulation the Maxwell equations on a generic (not necessarily flat) manifold and written in general coordinates. For a detailed introduction to the topic we refer to the relevant literature, e.g.\ \cite{Landau2,Post}. Throughout the whole paper natural units with $\epsilon_0 = \mu_0 = c = 1$ are used.

Greek indices $\mu, \nu, \rho, \ldots$ are spacetime indices and run from 0 to 3, Latin indices $i,j,k,\ldots$ space indices with values from 1 to 3. Furthermore an adapted coordinate system is used at the interface, such that $(x^i) = (x^A,x^\perp)$, where $x^A$ are the directions parallel to the interface, while $x^\perp$ is perpendicular. Therefore capital Latin indices take values 1,2.
 
For the metric we use the ``mostly plus'' convention, so the standard flat metric is $g_{\mu\nu} = \mbox{diag}(-1,1,1,1)$. Time is always interpreted as the zero-component of $x^{\mu}$, $x^0 = t$. With this identification an induced space metric can be obtained as \cite{Landau2}
\begin{equation}
\label{indmetric}
 \gamma^{ij} = g^{ij}\ , \qquad \gamma_{ij} = g_{lk} - \frac{g_{0i}g_{0j}}{g_{00}}\ , \qquad \gamma^{ij} \gamma_{jk} = \delta^i_k\ ,
\end{equation}
where $\delta^i_k$ is the Kronecker symbol.
This implies as relation between the determinant of the spacetime metric, $g$, and the one of the space metric, $\gamma$,
\begin{equation}
\label{detrel}
 -g = -g_{00} \gamma
\end{equation}
Furthermore we notice that under a coordinate transformation the spacetime metric transforms as
\begin{equation}
\label{dettransform}
 \bar g = \left \lvert \frac{\partial x^{\rho}}{\partial \bar x^{\mu}} g_{\rho \sigma} \frac{\partial x^{\sigma}}{\partial \bar x^{\nu}}\right \rvert = \left \lvert \frac{\partial x^{\rho}}{\partial \bar x^{\mu}}\right\rvert^2 g\ .
\end{equation}
In a similar way the relations
\begin{align}
\label{parallelmetric}
 \gamma^{AB}\left(\gamma_{BC}  - \frac{\gamma_{\perp B}\gamma_{\perp C}}{\gamma_{\perp \perp}}\right) &= \delta^A_C\ , \\
\label{parallelmetricII}
 \gamma_{AB}\left(\gamma^{BC}  - \frac{\gamma^{\perp B}\gamma^{\perp C}}{\gamma^{\perp \perp}}\right) &= \delta_A^C\ .
\end{align}
hold with respect to the adapted coordinate system. In this way an induced two-dimensional metric on the interface may be defined.

To motivate the transformation properties \eqref{barEB} and \eqref{barDH} we present the Maxwell equations in terms of the field strength tensor $F_{\mu\nu}$ and the excitation tensor $\mathcal H^{\mu\nu}$. The field strength tensor $F_{\mu\nu}$ encompasses the electric field and the magnetic induction, the excitation tensor $\mathcal H^{\mu\nu}$ the displacement vector and the magnetic field with the identification
\begin{align}
\label{spacevec1}
 E_i &= F_{0i}\ , & B^i &= - \frac{1}{2} \epsilon^{ijk} F_{jk}\ , \\
\label{spacevec2}
 D^{i} &= - \sqrt{-g_{00}} \mathcal H^{0i}\ , & H_i &= -\frac{\sqrt{-g_{00}}}{2} \epsilon_{ijk} \mathcal H^{jk}\ .
\end{align}
Finally, electric charge and current are combined into a four-current $J^{\mu} = (\rho/\sqrt{-g_{00}}, j^i/\sqrt{-g_{00}})$. In this way the Maxwell equations can be written in the compact form
\begin{align}
\label{EOMcomp}
 \epsilon^{\mu\nu\rho\sigma} \partial_\nu F_{\rho \sigma} &= 0\ , & D_\nu \mathcal H^{\mu \nu} &= - J^\mu\ .
\end{align}
The Maxwell equations depend on the metric through the covariant derivative $D_\mu$. Since
\begin{equation}
\label{covder}
 D_\nu \mathcal H^{\mu \nu} = (\partial_\mu + \Gamma^{\nu}_{\nu \rho}) \mathcal H^{\mu \rho} = \frac{1}{\sqrt{-g}} \partial_{\nu} (\sqrt{-g} \mathcal H^{\mu\nu})
\end{equation}
it is seen that the Maxwell equations just depend on the determinant of the metric, but not on its individual components.

Diffeomorphisms can change the orientation of a manifold, such that a right-handed coordinate system in laboratory space is mapped onto a left-handed one in electromagnetic space. This induces several changes of signs due to the Levi-Civita tensor that appears in the Maxwell equations. The four dimensional Levi-Civita tensor is defined as
\begin{align}
 \epsilon_{\mu\nu\rho\sigma} &= \sqrt{-g}[\mu\nu\rho\sigma]\ , & \epsilon^{\mu\nu\rho\sigma} &= - \frac1{\sqrt{-g}}[\mu\nu\rho\sigma]\ ,
\end{align}
with $[0123] = 1$. The relation between the four-dimensional Levi-Civita tensors in the three different spaces can be written as
\begin{equation}
 \epsilon_{\mu\nu\rho\sigma} = \bar s \frac{\sqrt{-g}}{\sqrt{-\bar g}} \bar \epsilon_{\mu\nu\rho\sigma} =  \bbar s \frac{\sqrt{-g}}{\sqrt{-\bbar g}} \bbar \epsilon_{\mu\nu\rho\sigma}\ ,
\end{equation}
where $\bar s$ and $\bbar s$ are $+1$ if the corresponding map does not change the orientation of the manifold, $-1$ otherwise

The reduction of the four dimensional to the three dimensional tensor reads
\begin{equation}
 \epsilon_{0ijk} = \sqrt{-g_{00}} \epsilon_{ijk}\ , \qquad \epsilon^{0ijk} = - \frac1{\sqrt{-g_{00}}} \epsilon^{ijk}\ .
\end{equation}
An additional complication arises in the definition of $\bar \epsilon_{ijk}$ and $\bbar \epsilon_{ijk}$, since the orientation of the spacetime manifold may change without changing the orientation of space (e.g.\ a map $\bar t = -t$, $\bar x^i = x^i$ changes the orientation of spacetime but not of space.) Therefore the corresponding relations should be written as
\begin{align}
 \bar \epsilon_{0ijk} &= \bar \sigma \sqrt{-\bar g_{00}} \bar \epsilon_{ijk}\ , & \bbar \epsilon_{0ijk} &= \bbar \sigma \sqrt{-\bbar g_{00}} \bbar \epsilon_{ijk}\ ,
\end{align}
where $\bar \sigma = +1$ if space and spacetime have the same orientation and $\bar \sigma = -1$ otherwise.

\providecommand{\href}[2]{#2}\begingroup\raggedright\endgroup

\end{document}